\documentclass[10pt,aps,pre,twocolumn,superscriptaddress,byrevtex,bibnotes,nofootinbib,showpacs,showkeys,a4paper,longbibliography]{revtex4-1}
\usepackage{microtype}
\usepackage{graphicx}
\usepackage{amsmath}
\usepackage{amsfonts}
\usepackage{amssymb}
\usepackage[colorlinks,linkcolor=blue,urlcolor=blue,citecolor=blue]{hyperref}
\usepackage{lineno}

\newcommand{\be}{\begin{equation}}
\newcommand{\ee}{\end{equation}}
\newcommand{\Esp}[1]{\left\langle #1 \right\rangle}
\newcommand{\Comment}[1]{}

\newcommand{\E}{\mathrm{e}}
\newcommand{\ra}{\rightarrow}
\newcommand{\reals}{\mathbb{R}}

\newcommand{\hlambda}{\hat\lambda}
\newcommand{\hG}{\hat G}
\newcommand{\hH}{\hat H}
\newcommand{\hI}{\hat I}
\newcommand{\ha}{\hat a}
\newcommand{\hp}{\hat p}
\newcommand{\hP}{\hat P}
\newcommand{\bx}{\bar x}
\newcommand{\bF}{\overline{F}}
\DeclareMathOperator{\var}{var}
\DeclareMathOperator{\stderr}{err}

\newcommand{\m}{M}
\renewcommand{\r}{R}
\newcommand{\fracpow}[3]{\left(\frac{#1}{#2}\right)^{#3}}

\begin{document}
\title[Convergence of large deviation estimators]{Convergence of large deviation estimators}

\author{Christian M. Rohwer}
\affiliation{Max-Planck-Institut f\"ur Intelligente Systeme, Heisenbergstra\ss e 3, D-70569 Stuttgart, Germany}
\affiliation{Institut f\"ur Theoretische Physik IV, Universit\"at Stuttgart, Pfaffenwaldring 57, D-70569 Stuttgart, Germany}
\affiliation{Department of Physics and Institute of Theoretical Physics, Stellenbosch University, Stellenbosch 7600, South Africa}

\author{Florian Angeletti}
\affiliation{National Institute for Theoretical Physics (NITheP), Stellenbosch 7600, South Africa}

\author{Hugo Touchette}
\email{htouchette@sun.ac.za}
\affiliation{National Institute for Theoretical Physics (NITheP), Stellenbosch 7600, South Africa}
\affiliation{Department of Physics and Institute of Theoretical Physics, Stellenbosch University, Stellenbosch 7600, South Africa}

\date{\today}

\begin{abstract}
We study the convergence of statistical estimators used in the estimation of large deviation functions describing the fluctuations of equilibrium, nonequilibrium, and manmade stochastic systems. We give conditions for the convergence of these estimators with sample size, based on the boundedness or unboundedness of the quantity sampled, and discuss how statistical errors should be defined in different parts of the convergence region. Our results shed light on previous reports of `phase transitions' in the statistics of free energy estimators and establish a general framework for reliably estimating large deviation functions from simulation and experimental data and identifying parameter regions where this estimation converges.
\end{abstract}



\maketitle

\section{Introduction}

The generating function of a fluctuating quantity or random variable $X$, defined as
\be
G(k)=\Esp{\E^{k X}}=\int_{-\infty}^\infty  p(x)\, \E^{k x}\,dx,
\ee 
where $p(x)$ is the distribution of $X$, is related to many important physical quantities. Examples include velocity fields of turbulent fluids, whose generating function, estimated in terms of moments, is related to energy dissipation and multifractal exponents \cite{mccauley1990,mandelbrot1999,harte2001}, the generating function of the energy of systems at thermal equilibrium, which is essentially the partition function \cite{lanford1973,ellis1985,oono1989,touchette2009}, and the generating function of the work performed on nonequilibrium systems, which is linked via Jarzynski's equality to equilibrium free energy differences \cite{jarzynski1997}. In large deviation theory, generating functions are also related to so-called rate functions, which give the likelihood of rare fluctuations in stochastic systems. In recent years, this theory has been applied successfully to describe the fluctuations of equilibrium systems \cite{lanford1973,ellis1985,oono1989,touchette2009} and nonequilibrium systems driven in steady states \cite{hollander2000,derrida2007,bertini2007,harris2013}, in addition to manmade systems such as queues and networks \cite{shwartz1995,dembo1998,montanari2002,kishore2012,bacco2015}.

The problem that we consider in this paper is the statistical estimation of generating functions from empirical data which arises experimentally or numerically in all the applications mentioned above. To be more precise, we consider a finite sample $\{X^{(j)}\}_{j=1}^M$ of $M$ realizations (copies, repetitions or empirical values) of the random variable $X$ and study the convergence of the following statistical estimator of $G(k)$:
\be
\hG_M(k)=\frac{1}{M}\sum_{j=1}^M \E^{kX^{(j)}}.
\label{eqest0}
\ee
This function converges pointwise to $G(k)$ as the sample size $M$ increases, but a major problem is that it does not converge \emph{uniformly} in $k$, which means that the sample size $M$ needed to achieve a given accuracy for $\hG_M(k)$ depends on $k$. In fact, it is known that, depending on the random variable considered, $\hG_M(k)$ converges slowly for some $k$ and, in some cases, does not converge at all. This problem, often referred to as the `linearization problem', has been studied in the context of multifractal analysis \cite{abry2007,muzy2008,bacry2010} and glassy phase transitions \cite{derrida1981,bouchaud1997,berthier2011}. Convergence problems have also been studied for the so-called Jarzynski estimator, which is an estimator similar to (\ref{eqest0}) used to obtain free energy differences from nonequilibrium experiments~\cite{hummer2001,liphardt2002,harris2007b,hummer2010,gupta2011,alemany2012,kim2012}. The focus of these studies, however, is mostly on the statistical bias of $\hG_M(k)$ \cite{wood1991,gore2003,zuckerman2002,zuckerman2002b,palassini2011,suarez2012}, which disappears in the limit $M\ra\infty$, rather than the convergence of $\hG_M(k)$ as a function of $k$ and $M$.

In this paper, we study this convergence for estimating large deviation functions. Our starting point is a series of studies on data networks \cite{crosby1997,duffield1995,lewis1998,duffy2005} showing that large deviation functions can be estimated efficiently from generating functions for random variables having a finite number of values, such as random bits, and for bounded random variables, such as uniform variates. Here, we extend these studies by considering \emph{unbounded} random variables, which naturally arise when considering observables of equilibrium and nonequilibrium systems. For these, we show that the estimation of large deviation functions based on $\hG_M(k)$ converges up to some critical value $k_c$, which depends on $M$ and the tail of the observable distribution. Moreover, we show that standard statistical errors for this estimator can be defined only up to $k_c/2$. The knowledge of $k_c$ is thus important for properly evaluating, for a given sample size, the parameter range for which large deviation functions are estimated reliably.

These functions play an important role, as mentioned, for characterizing the typical states and fluctuations of equilibrium, nonequilibrium, and manmade stochastic systems. The numerical computation of these functions for observables of nonequilibrium systems (e.g., particle and energy currents, work, heat, activity, entropy production) has been the subject of active studies in the last years (see, e.g., \cite{lecomte2007a,tailleur2007b,giardina2011,gorissen2009,gorissen2011,merolle2005,hedges2009,chandler2010,hurtado2014}), contributing to our understanding of nonequilibrium phase transitions and fluctuation symmetries. More recent works are now attempting to estimate rate functions in real experiments, for example, in active-matter systems \cite{kumar2011,kumar2015}. On the experimental side, large deviation estimations have also been carried out, as mentioned, for data networks and provide in this context a real-time estimate of overflow probabilities in data servers~\cite{crosby1997,duffield1995,lewis1998,duffy2005}.

Our results provide for these applications a general method for estimating large deviation functions, their errors, and their convergence region from finite data samples. They can also be applied for computing multifractal spectra of random fields or time series, dispersion exponents in sheared flows \cite{haynes2014}, in addition to free energy differences from nonequilibrium work experiments \cite{hummer2001,liphardt2002,harris2007b,hummer2010,gupta2011,alemany2012,kim2012}. Conceptually, these problems all fall in the topic of large deviation estimation.

The outline of the paper is as follows. The large deviation estimators that we consider are defined in Sec.~\ref{secres}. Test cases involving bounded and unbounded random variables are considered in Sec.~\ref{sectest} to show how the estimators' convergence region depends on sample size, and how this dependence can be determined, a priori, from the knowledge of the distribution considered or, a posteriori, from a sample of that distribution. Most of our results are illustrated for simplicity for sums of independent and identically distributed random variables. In Sec.~\ref{secdisc} we discuss applications for correlated Markov processes and systems composed of many interacting particles, for which the distributions of observables typically scale with time and the number of particles, respectively. Final conclusions are given in Sec.~\ref{seccon}.

\section{Method and results}
\label{secres}

\subsection{Estimators}

The theory of large deviations is concerned with random variables $A_n$, depending on some parameter $n$, whose probability distribution $p_n(a)=P(A_n=a)$ decays approximately exponentially as
\be
p_n(a)\approx \E^{-n I(a)}
\label{eqldp1}
\ee
when $n\ra\infty$, with sub-exponential corrections in $n$ \cite{dembo1998,hollander2000,touchette2009}. This approximation appears naturally in many equilibrium and nonequilibrium systems, where $A_n$  represent, for example, the energy or magnetization of $N$ particles occupying a volume $V$, in which case $n=N$ or $n=V$ \cite{ellis1985,oono1989,touchette2009}, or the current or heat integrated over a time $T$, so that $n=T$ \cite{derrida2007,bertini2007,harris2013}. In manmade systems, $A_n$ can also be the number of `customers' served in a queue \cite{shwartz1995} or the fraction of time spent by a random walker on specific sites of a network after $n$ time steps \cite{montanari2002,kishore2012,bacco2015}. 
In all cases, the distribution of $A_n$ is completely characterized to leading order in $n$ by the function $I(a)$ which gives the likelihood of small and large fluctuations of $A_n$ around its equilibrium or stationary value. This function is called the \emph{rate function} in large deviation theory \cite{ellis1985} and has the interpretation in physics of an entropy function or a generalized potential, depending on the application considered \cite{touchette2009}.

Many techniques can be used to find $I(a)$. The most common proceeds by calculating the so-called \emph{scaled cumulant generating function} (SCGF), defined as 
\be
\lambda(k)=\lim_{n\ra\infty} \frac{1}{n}\ln \Esp{ \E^{ n k A_n} }.
\label{eqgenscgf1}
\ee
Provided that this limit function exists for $k\in\reals$ and is differentiable, it is known that $p_n(a)$ has the large deviation form of (\ref{eqldp1}) and that its rate function is given by the Legendre transform of $\lambda(k)$:
\be
I(a)=k_a a-\lambda(k_a),
\label{eqlt1}
\ee
$k_a$ being the root of $\lambda'(k)=a$ \cite{dembo1998,hollander2000,touchette2009}.\footnote{This holds for convex rate functions. For results on nonconvex rate functions, see Sec.~4.4 of \cite{touchette2009}, and \cite{touchette2010b}.} Consider, for example, the case where $A_n$ is a sample mean of $n$ independent and identically distributed (IID) random variables:
\be
A_n=\frac{1}{n}\sum_{i=1}^n X_i.
\label{eqsm1}
\ee
Then the SCGF takes the simple form
\be
\lambda(k)=\ln \langle \E^{kX}\rangle,
\label{simplescgf1}
\ee
so that the large deviation rate function is obtained as the Legendre transform of the cumulant function of a single random variable, denoted above by $X$ without the subscript because of the IID property. For other observables $A_n$ involving \emph{correlated} random variables, the calculation of $I(a)$ is more involved, but still proceeds from $\lambda(k)$ as defined in (\ref{eqgenscgf1}).

In many applications, the SCGF must be estimated empirically from data samples. For the IID sample mean (\ref{eqsm1}), to take the simplest example, this estimation proceeds by accumulating a sample $\{X^{(j)}\}_{j=1}^\m$ of $\m$ IID copies of the random variable $X$, distributed according to the (a priori unknown) distribution $p(x)$, and by approximating the generating function of $X$ by the estimator $\hG_M(k)$ as defined in (\ref{eqest0}). The estimator of $\lambda(k)$ is then defined as \cite{touchette2011}
\be
\hlambda_\m(k)=\ln \hG_\m(k).
\label{eqest1}
\ee
Our goal in this paper is to understand the convergence of this estimator as a function of $\m$ and $k$. From now on, we consider the IID case to simplify the discussion; the case of correlated random variables and observables other than sums is discussed in Sec.~\ref{secdisc}. 

The estimator of rate functions that we consider is defined from the Legendre transform (\ref{eqlt1}) by noting that the estimator (\ref{eqest1}) of the SCGF is necessarily analytic, since it is a finite sum of exponentials, and is thus differentiable for all $\m<\infty$. As a result, we consider 
\be
\hI_\m(a)=k_a a-\hlambda_\m(k_a),
\label{eqesti1}
\ee
as an estimator of $I(a)$, where $k_a$ is the computed root of $\hlambda_\m'(k)=a$ \cite{touchette2011}. Alternatively, we can proceed parametrically by fixing $k$, and obtain $I$ at the estimated value
\be
\ha_\m(k)=\lambda'_\m(k)=\frac{\sum_{j=1}^\m X^{(j)}\E ^{k X^{(j)}}}{\sum_{j=1}^\m \E^{k X^{(j)}}}
\label{eqha1}
\ee
using
\be
\hI_\m(\ha_\m)=k \ha_\m-\hlambda_\m(k).
\label{eqesti2}
\ee
Strictly speaking, the estimators (\ref{eqesti1}) and (\ref{eqesti2}) are different. We have found in all of our numerical tests, however, that they are nearly identical and differ only because of the discretization used for $k$. This is a minor, non-statistical source of errors, which is not discussed further. 

As statistical estimators, $\hlambda_\m(k)$ and $\hI_\m(a)$ converge pointwise to $\lambda(k)$ and $I(a)$, respectively, in the limit of infinite sample size $\m\ra\infty$. Their speed of convergence was studied in \cite{duffy2005}, following previous results on overflow probabilities and bandwidth estimates of data networks \cite{crosby1997,duffield1995,lewis1998,duffy2005}. These studies, however, consider only bounded random variables for which $\hlambda_\m(k)$ and $\hI_\m(a)$ are known to converge quickly and uniformly. In this case, the probability distribution of both estimators has the large deviation form of (\ref{eqldp1}), which implies that these estimators converge exponentially fast for all $k$ or $a$ with $\m$ \cite{duffy2005}.

We extend these results in what follows by considering \emph{unbounded} random variables. In this case, the convergence of $\hG_\m(k)$ is limited by two problems, namely: the linearization effect, which leads to noisy tails of $\hG_\m(k)$, and the non-uniform convergence of $\hG_\m(k)$ in $k$, which means that its statistical error is not uniform in $k$. These problems are explained next and lead us to define, as mentioned, a threshold value $k_c$ depending on $\m$ for the  convergence of $\hG_\m(k)$, $\hlambda_\m(k)$, and $\hI_\m(a)$. Applications of these results are presented in the next section.

\subsection{Linearization effect}

The linearization effect refers to the fact that sums of exponentials, such as in (\ref{eqest0}), are dominated as $k\ra\infty$ by the largest sample element
\be
X_{\max}=\max_{1\leq j\leq \m}  X^{(j)},
\ee
so that
\be
\sum_{j=1}^\m \E^{kX^{(j)}} \approx \E^{kX_{\max}},\qquad k\ra\infty.
\ee
In this regime, the SCGF estimator thus becomes linear in $k$:
\be
\hlambda_\m(k)\approx k X_{\max},\qquad k\ra\infty.
\ee
Similarly, for $k\ra -\infty$, the sum is dominated by the smallest element
\be
X_{\min}=\min_{1\leq j\leq \m} X^{(j)},
\ee
so that
\be
\hlambda_\m(k)\approx k X_{\min},\qquad k\ra -\infty.
\ee

If the sample $\{X^{(j)}\}$ is obtained from a discrete or continuous distribution with \emph{bounded} support, then the values of $X_{\max}$ and $X_{\min}$ are also bounded and the linearization effect is real: that is, the asymptotic linear branches of $\hlambda_\m(k)$ seen for $|k|\ra \infty$ correspond in this case to actual linear branches of $\lambda(k)$ and are unlikely to change much as the sample size $\m$ is increased, since the sample will most likely `fill' the range of the bounded distribution for $\m$ large enough. However, if the sample is obtained from an \emph{unbounded} distribution, then the linear tails of $\hlambda_\m(k)$ coming from $X_{\max}$ and $X_{\min}$ are an artifact of the finite-size sample: $X_{\max}$ and $X_{\min}$ fluctuate from sample to sample, which implies that $\hlambda_\m(k)$ has fluctuating linear tails for large $|k|$ which are not related to the actual tails of $\lambda(k)$. 

This problem affects not only large deviation computations, as mentioned in the introduction: any estimation of exponential sums, such as those entering in   free energy computations and multifractal analysis \cite{abry2007,muzy2008,bacry2010}, is limited by the linearization effect whenever unbounded random variables are considered. The main problem in these cases is to identify the \emph{onset of linearization}, that is, the threshold value $k_c$ such that, for $|k|<k_c$, $\hlambda_\m(k)$ is not affected artificially by linearization and is therefore a good representation of $\lambda(k)$. 

In general, $k_c$ depends on $\m$ as well as the particular distribution considered. Moreover, for asymmetric distributions, two threshold values must be considered: $k_c^{-}$ for the left tail of $p(x)$ determining the distribution of $X_{\min}$, and $k_c^{+}$ for the right tail of $p(x)$ determining $X_{\max}$. The convergence and errors of estimators thus depend on whether $k\in [k_c^{-},k_c^{+}]$.

In general, it is difficult to determine $k_c^{-}$ and $k_c^+$ exactly; for practical purposes, however, it is sufficient to approximate their growth as $\m\ra\infty$. This can be done by approximating $G(k)$ in the limit $k\ra\infty$ using the saddle-point or Laplace approximation \cite{bender1978} as
\be
G(k)\approx \E^{k x^*(k)+\ln p(x^*(k))},
\label{eqlaplace1}
\ee
where $x^*(k)$ satisfies
\be
k p(x^*)+p'(x^*)=0.
\ee 
This shows that $G(k)$ is determined for large $k$ by a narrow region of the distribution $p(x)$ located around the \emph{saddle} or \textit{concentration point} $x^*(k)$ \cite{angeletti2011,touchette2005,jarzynski2006}.\footnote{Including the Gaussian correction to the saddle-point only leads to subdominant corrections to the scaling of $k_c$ with $M$.} As a result, $\hG_\m(k)$ must be a good estimator of $G(k)$ when the empirical distribution or density histogram $\hp_\m(x)$ of the sample $\{X^{(j)}\}_{j=1}^\m$ is close to $p(x)$ around $x^*(k)$.

To express this more quantitatively, we define a typicality region for the random variable $X$ by considering the probability
\be
P(X^{(1)},\ldots,X^{(\m)}<\bx)=P(X_{\max}< \bx)
\ee
that all the sample elements $X^{(j)}$ are smaller than a given bound $\bx$.\footnote{We could also consider only a fraction of the $X^{(j)}$'s to be below $\bx$; however, this does not significantly alter the scaling of $k_c$.} This probability is given in terms of the cumulative distribution
\be
F(x)=P(X< x)=\int_{-\infty}^x p(z)\, dz
\ee
by
\be
P(X^{(1)},\ldots,X^{(\m)}<\bx)=F(\bx)^\m,
\label{eqprobreal1}
\ee
and can be approximated for $\bx$ and $\m$ large by
\be
P(X^{(1)},\ldots,X^{(\m)}<\bx)\approx 1- \m\bF(\bx),
\label{eqprob1}
\ee
where $\bF(x)=1-F(x)$ is the complementary cumulative distribution of $X$. From this, we see that, if $\bx$ is a constant independent of $\m$, then the probability (\ref{eqprobreal1}) vanishes as $\m\ra\infty$, as all the samples eventually reach $\bx$. However, if we scale $\bx$ as a function of $\m$, then the same probability will in general not go to zero. In particular, if we set
\be 
\bx=\bx(\m,\tau)=\bF^{-1}\left(\frac{\tau}{\m}\right),
\label{eq:bx:def}
\ee
where $\tau$ is an arbitrary small constant and $\bF^{-1}$ is the inverse of $\bF$, then the probability of having all the samples smaller than $\bx$ is equal to $\E^{-\tau}$ for all $\m$. The region $(-\infty,\bx]$ therefore defines a typical region for the sample $\{X^{(j)}\}$: as $\m$ grows, all samples will fall in that region with constant probability.

With this result we now define the truncated generating function
\be
G_{\m,\tau}(k)=\int_{-\infty}^{\bx(\m,\tau)} \E^{kx}\, p(x)\, dx. 
\ee
Depending on $k$ and $\m$, two different situations then arise for $x^*(k)$. On the one hand, if $x^*(k)< \bx(\m,\tau)$, then
\be
\hG_\m(k)\approx G_{\m,\tau}(k)\approx G(k),
\ee
and we are away from the linearization regime. On the other hand, if $x^*(k)>\bx(\m,\tau)$, then $G(k)$ is not well approximated by $G_{\m,\tau}(k)$ or $\hG_\m(k)$ since $x^*(k)$, the concentration point of $G(k)$, lies outside the typical values covered by the sample. In this case, one must either increase $\m$ for a given $k$ so that $\bx(\m,\tau)$ reaches $x^*(k)$, or decrease $k$ for a given $\m$ so that $x^*(k)$ reaches $\bx(\m,\tau)$. The threshold value of $k$ for which $x^*=\bx$ defines $k_c$; thus,
\be
x^*(k_c)=\bx(\m,\tau).
\label{eqkcest1}
\ee
This result yields the upper bound $k_c^+$; a similar calculation yields the lower bound $k_c^-$ mentioned before by considering $P(X_{\min}>\bx)$ instead of $P(X_{\max}<\bx)$.

This analysis gives estimates for $k_c^-$ and $k_c^+$ that are good enough for practical purposes, as they capture the scaling of the linearization effect with $\m$ based on the tail behavior of $p(x)$ in (\ref{eqlaplace1}).\footnote{A similar analysis of sample extremes was developed for a more specific model by Hurtado and Garrido \cite{hurtado2009b} to study statistical errors in the cloning algorithm \cite{lecomte2007a,tailleur2007b,giardina2011}.} For example, if $X$ is distributed according to the Gaussian density with $p(x)\sim \E^{-x^2/2}$ as $|x|\ra\infty$, then we obtain from \eqref{eq:bx:def}
\be
\bx (M, \tau) =  2\, \mathrm{erfc}^{-1}\left(\frac 2 \m \right).
\label{eq:ex:gauss:xb}
\ee
Moreover, the concentration point for this distribution is located at $x^*(k)= k$. Combining this with \eqref{eq:ex:gauss:xb} in \eqref{eqkcest1} and using known asymptotics for the complementary error function then yields
\be
k_c^{\pm}\approx\pm \sqrt{\ln \m}.
\label{eqestg1}
\ee
More generally, if 
\be
p(x)\approx \E^{-|x|^\rho},\qquad \rho>1,
\label{eqexpdist1}
\ee
as $|x|\ra\infty$, then (\ref{eqkcest1}) yields 
\be
k_c^{\pm}\approx\pm (\ln \m)^{1-1/\rho}.
\ee
The full derivation of this result can be found in \cite{angeletti2012}. For both cases, the estimate of $k_c$ does not depend on $\tau$, as shown in \cite{angeletti2012}. If, however, $p(x)$ is bounded from above at $x_{\max}$ and behaves like
\be
p(x)\approx (x_{\max}-x)^{\beta},\qquad \beta>0,
\label{eqpest1}
\ee
for $x<x_{\max}$ as $x\ra x_{\max}$ from below, then
\be
k_c^{+}\approx \left(\frac{\m}{\tau}\right)^{\frac{1}{\beta+1}}.
\label{eqkcest2}
\ee
In this case, there is an explicit dependence on $\tau$, which for applications can be set to some fixed but otherwise arbitrary value. A similar scaling is obtained for bounded (e.g., uniform) random variables and finite, discrete random variables.

\subsection{Statistical errors}

Estimators must be supplemented by statistical errors to be meaningful. Commonly, this is done by assuming that the distribution of the sum defining an estimator converges to a Gaussian distribution around its mean, which leads to defining the `dispersion' or error of the estimator as its standard error. In our case, we have to be careful with this error definition: since the variance of the random variable $\E^{kX}$ is
\be
\var(\E^{kX})=G(2k)-G(k)^2,
\ee
the variance of $\hG_\m(k)$ is defined only on half the range on which $\hG_\m(k)$ converges. Moreover, although that estimator is known to converge for $k<k_c$, that convergence may not be to a Gaussian random variable, which prevents us from using the standard deviation as an error measure already from $k_c/2$.

This basic observation is supported by rigorous mathematical results obtained recently by Ben Arous and collaborators \cite{arous2005} which show for a general class of random variables\footnote{This class includes the Gaussian distribution, the Gamma distribution, and the stretched exponential distribution (\ref{eqexpdist1}) among many others.} that $\hG_\m(k)$ converges, when properly rescaled, to a Gaussian random variable for all $k=k_c(\m)/\alpha$ when $\alpha>2$. It then converges to a L\'evy $\alpha$-stable random variable for $k=k_c(\m)/\alpha$ when $1<\alpha\leq 2$, whereas there is no convergence when $\alpha<1$. This means overall that we have to consider three regions for defining error bars:
\begin{enumerate}
\item $k\leq k_c(\m)/2$: $\hG_\m(k)$ is \emph{self-averaging}, meaning that it converges in probability to $G(k)$ as $\m\ra\infty$. Moreover, this estimator is  Gaussian-distributed around $G(k)$, so that its standard variance can be used as an error estimate;
\item $k_c(\m)/2<k\leq k_c(\m)$: $\hG_\m(k)$ is self-averaging, but is not Gaussian-distributed around $G(k)$, so that standard (Gaussian) error bars are inadequate;
\item $k>k_c(\m)$: $\hG_\m(k)$ is not self-averaging, so there is no convergence to $G(k)$ as $\m\ra\infty$.
\end{enumerate}
We detail each region next and explain its consequences for defining errors for $\hG_\m(k)$, $\hlambda_\m(k)$ and $\hI_\m(a)$. For simplicity, we only discuss the upper bound $k_c^+$; errors concerning $k_c^-$ are defined similarly.

\subsubsection{Error estimates below $k_c/2$}

In this region, the error bar for $\hG_\m(k)$ can be defined as its standard deviation, which is estimated from the empirical variance:
\be
\stderr[\hG_\m(k)] =\sqrt{\var[\hG_\m(k)]}
=\frac{1}{\sqrt{\m}}\sqrt{\hG_\m(2k)-\hG_\m(k)^2}.
\ee
Computing from this error the error of $\hlambda_\m(k)$ is not trivial because the latter is a nonlinear function of $\hG_M(k)$. However, for small errors we can perform a Taylor expansion of (\ref{eqest1}) to first order, as commonly done in physics \cite{pohorille2010}, to obtain
\be
\stderr[\hlambda_\m(k)]=\frac{\stderr[\hG_\m(k)]}{\hG_\m(k)}.
\ee

With this error, we can define the error of the rate function estimator $\hI_\m(a)$ parametrically by varying $k$, as explained before. For a given $k<k_c(\m)/2$, we first compute $\ha_\m(k)$ as in (\ref{eqha1}). Denoting the numerator of the right-hand side of (\ref{eqha1}) by $\hH_\m(k)$, we next estimate the error of $\ha_\m(k)$ as
\be
\stderr[\ha_\m(k)]=\sqrt{\fracpow{\stderr[\hH_\m(k)]}{\hG_\m(k)}{2} + \fracpow{ \stderr[\hG_\m(k)] \hH_\m(k)}{\hG_\m(k)^2}{2}},
\label{eqerra1}
\ee
which follows by applying a Taylor expansion to the definition of $\ha_\m(k)$ and by neglecting the correlation between the numerator and denominator.\footnote{The numerator and denominator of (\ref{eqha1}) are not independent, but this is a necessary approximation to be able to obtain an error estimate.} Given the Legendre transform (\ref{eqlt1}) or (\ref{eqesti1}), it is then natural to define the error for $\hI_\m(a)$ at $a=\ha_\m(k)$ as the sum of the errors:
\be
\stderr[\hI_\m(\ha_\m(k))]= \sqrt{ k^2\stderr[\ha_\m(k)]^2 + \stderr[\hlambda_\m(k)]^2}.
\label{eqerri1}
\ee
This quantity probably overestimates the error; however, it is the simplest error that one can define, based on the linear form of the Legendre transform, which does not assume any dependency between $\ha_\m$ and $\hlambda_\m$. 

\subsubsection{Error estimates between $k_c/2$ and $k_c$}

In this region, linearization sets in from $k_c(\m)$, leading $\hG_\m(k)$ to converge to $G(k)$ but in a non-Gaussian way, which implies that the standard deviation calculated from $M$ realizations cannot be used as an error estimate. In this case, it is common to define the error of estimators not from one sample $\{X^{(j)}\}_{j=1}^\m$, but from $\r$ such samples of size $\m$, called \textit{repetitions}. For the SCGF this means generating $\r$ independent samples of size $\m$ yielding $\r$ estimators $\hlambda_\m^{(\ell)}(k)$, $\ell=1,\ldots,\r$, which are averaged to yield the following estimate of $\lambda(k)$:
\be
\hlambda_{\r\times \m}(k)=\frac{1}{\r}\sum_{\ell=1}^\r \lambda_\m^{(\ell)}(k).
\ee
The error for this estimation is then obtained by computing the standard error over the $\r$ repetitions:
\be
\stderr[\hlambda_{\r\times \m}(k)]=\frac{ \stderr[\hlambda_{\m}(k)] } {\sqrt \r}
\label{eqerrest2}
\ee
The error estimate for $\hI_\m(a)$ can be defined similarly using repetitions and the error method presented for one sample. In this case, the repetition error $\stderr[\ha_{\r\times \m}(k)]$ must be computed as in (\ref{eqerrest2}) and added as in (\ref{eqerri1}) to the repetition error of $\hlambda_{\r\times \m}(k)$. 

Though more computationally intensive, the use of repetitions provides better error estimates for $\hlambda_\m$ and $\hI_\m$, as the logarithm in $\hlambda_\m(k)$ has the effect of regularizing the extreme values (and thus the linearization) of $\hG_\m(k)$. In practice, a sufficiently large sample can be partitioned into $\r$ smaller samples to apply this method. Alternatively, bootstrap methods can be used to generate new samples from the empirical distribution of the sample already obtained \cite{efron1979,efron1994,davison1997}.


\subsubsection{Error estimates above $k_c$}

In this region, estimators do not converge, leaving the computation of $\hG_\m(k)$, $\hlambda_\m(k)$, and $\hI_\m(a)$ meaningless. To have an idea of the variability of these estimators, one can estimate them over $\r$ repetitions involving $\m$ samples, as before, and extract the first decile and last decile of these realizations. This can be taken as a measure of the error. Our results indicate, however, that such an error is typically very large and only confirms the fact that no useful information can be inferred about $\lambda(k)$ beyond $k_c(\m)$.

\section{Test cases}
\label{sectest}

We illustrate in this section the previous results about estimator convergence for four types of distributions: Gaussian, exponential, Bernoulli, and power-law. Gaussian distributions have been extensively studied in the context of the Jarzynski estimator \cite{wood1991,gore2003,zuckerman2002,zuckerman2002b,palassini2011,suarez2012} and are revisited here to illustrate the case of unbounded random variables. The exponential distribution is considered as a limiting case of the saddle-point analysis, whereas Bernoulli random variables illustrate our results for the bounded case and are relevant for data network applications \cite{crosby1997,duffield1995,lewis1998,duffy2005}. Finally,  power-law distributions are considered to discuss the case where $\lambda(k)$ diverges and large deviation functions do not exist. Other distributions fall, as will be explained, in each of these cases with only minor differences in the behavior of $\hlambda_\m(k)$ and $k_c$. Physical applications and non-IID random variables are discussed in the next section.

\subsection{Unbounded distributions}

\begin{figure*}[t]
\centering
\includegraphics{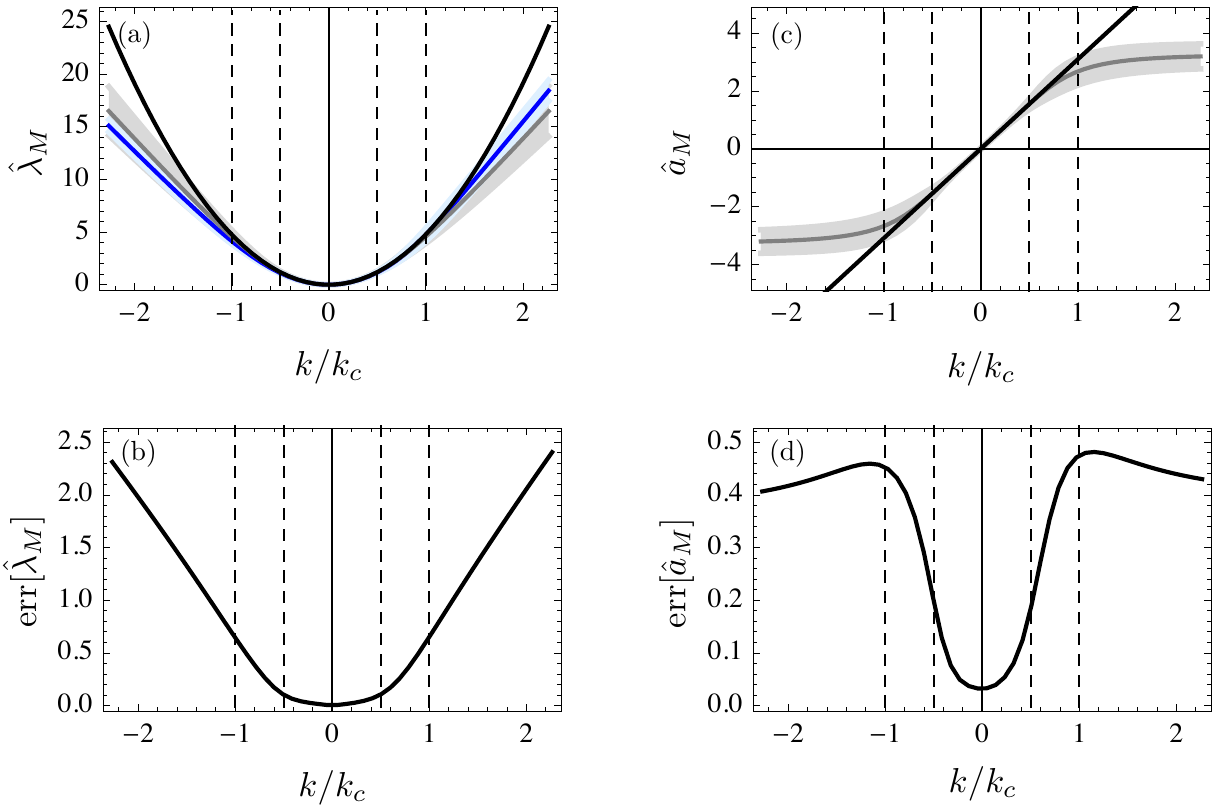}
\caption{(Color online) SCGF estimator for Gaussian random variables. Parameters: $\mu=0$, $\sigma=1$, $\m=1000$, and $\r=200$ repetitions (except otherwise noted). (a) Blue curve: $\hlambda_\m(k)$ with statistical error (blue shaded area) for a single repetition ($R=1$). Gray curve: $\hlambda_\m(k)$ with statistical error (gray shaded area). Black curve: Exact $\lambda(k)$. (b) Statistical error of $\hlambda_\m(k)$ showing the linearization effect. (c) Estimator $\ha_\m(k)$ of the derivative of $\lambda(k)$ with statistical error (gray shaded area). (d) Statistical error for $\ha_\m(k)$.}
\label{figgauss1}
\end{figure*}

We consider as in (\ref{eqsm1}) a sample mean $A_n$ of $n$ IID random variables and assume that these are distributed according to a Gaussian distribution with mean $\mu$ and variance $\sigma^2$. The exact SCGF in this case is
\be
\lambda(k)=\mu k+\frac{\sigma^2}{2} k^2,\qquad k\in\reals.
\label{eqscgfgauss1}
\ee

Figure~\ref{figgauss1}(a) shows the estimation of this SCGF based on the estimator $\hlambda_\m(k)$ using a sample $\{X^{(j)}\}_{j=1}^\m$ of $\m=1000$ Gaussian random variables with $\mu=0$ and $\sigma^2=1$. The relatively low sample size is used to obtain visible error bars. Moreover, rather than plotting $\hlambda_\m$ as a function of $k$ and identifying $k_c$ for varying $\m$, we plot the estimator as a function of $k/k_c$ using the priori estimate given in (\ref{eqestg1}), so that the convergence region is fixed at $|k/k_c|=1$ for all $\m$. In this way, plots of $\hlambda_\m$ obtained for different $\m$ look similar, provided that $\m$ is large enough; hence we show only one value of $\m$ in Fig.~\ref{figgauss1}(a). Note that, because of the choice $\mu=0$, we have $-k_c^-=k_c^+=k_c$; for asymmetric Gaussian distributions, $|k_c^-|$ is slightly different from $k_c^+$, but this does not affect the scaling of $k_c$.

The convergence and linearization regions of $\hlambda_\m(k)$ are clearly visible in Fig.~\ref{figgauss1}(a). For $|k/k_c|<1$, $\hlambda_\m(k)$ agrees with the exact $\lambda(k)$ of (\ref{eqscgfgauss1}) within the statistical errors determined from either $\r=1$ or $\r>1$ repetitions. For $|k/k_c|<1/2$, we have checked that both errors have similar magnitude (not shown), whereas for $1/2<|k/k_c|<1$ the two errors differ slightly (also not shown). More importantly, for $|k/k_c|>1$, $\hlambda_\m(k)$ starts to differ significantly from the exact $\lambda(k)$ because of the linearization coming from the maximum and minimum sample values. The gray curve in Fig.~\ref{figgauss1}(a) shows this linearization for the $\r$ repetition estimate as compared to the single repetition (blue curve). The former is in general more stable than the latter because of the averaging coming from the $\r$ repetitions; however, both estimators give results that are off the exact SCGF because $\hlambda_\m(k)$ and its error do not converge for $|k/k_c|>1$.

The linearization effect is also seen in the repetition error of $\hlambda_\m$ [Fig.~\ref{figgauss1}(b)] and the estimator $\ha_\m(k)$ [Fig.~\ref{figgauss1}(c)] of the derivative of $\lambda(k)$. Linearization appears for $\ha_\m(k)$ as plateaus with heights given in the $\r$ repetition case by the mean of the different maxima and minima contained in the repeated samples. Since the variance of these minima and maxima is independent of $k$, the statistical error of $\ha_\m$ is constant, as seen in Figs.~\ref{figgauss1}(c) and \ref{figgauss1}(d). Inside the convergence region, $|k/k_c|<1$, $\stderr[\ha_\m]$ decreases sharply from $|k/k_c|=1$ to $|k/k_c|=1/2$, below which it converges to 0 as $\m\ra\infty$ for any $\r\geq1$. This error behavior is interesting for two reasons. First, it can be interpreted as a `phase transition' or crossover as $k$ is varied, reflecting the transition from Gaussian to non-Gaussian errors at $|k/k_c|=1/2$. A similar crossover was reported in the behavior of the bias of the Jarzynski estimator \cite{suarez2012} and the partition function of glassy systems \cite{angeletti2011}. Second, it provides a simple way of estimating $k_c$ numerically without knowing the distribution of the random variables considered: we simply have to find the function $k(\m)$ that aligns the maximum of $\stderr[\ha_\m(k)]$ for different $\m$.

\begin{figure}[t]
\centering
\includegraphics{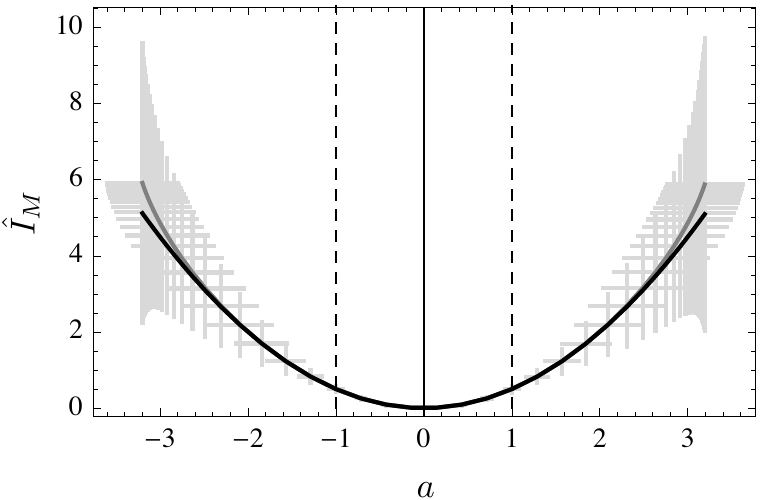}
\caption{Rate function estimator for Gaussian random variables. Parameters: $\mu=0$, $\sigma=1$, $\m=1000$, and $\r=200$. Gray curve: $\hI_\m(a)$ with statistical error bars. Black curve: Exact rate function.}
\label{figgauss2}
\end{figure}

From the estimators $\hlambda_\m$ and $\ha_\m$ we may now estimate the rate function $I(a)$ using the parametric relations (\ref{eqha1}) and (\ref{eqesti2}). The result is shown in Fig.~\ref{figgauss2} together with the exact result
\be
I(a)=\frac{(a-\mu)^2}{2\sigma^2},\qquad a\in\reals.
\ee
We show in this plot the vertical error bars for the ordinate $\hI_\m(\ha_\m(k))$ obtained from (\ref{eqerri1}), as well as horizontal error bars for the abscissa $\ha_\m(k)$ corresponding to the repetition error $\stderr[\ha_\m(k)]$ given in (\ref{eqerra1}). Also indicated is the value $\ha_\m(k_c)$, corresponding for $\mu=0$ and $\sigma=1$ to $a=1$, which bounds the convergence region of $I(a)$ where the error bars decrease as $\r$ and $\m$ are increased. In the results shown in Fig.~\ref{figgauss2}, the errors for $|a/\ha_\m(k_c)|<1$ are actually smaller than the width of the curves, whereas they increase substantially for $|a/\ha_\m(k_c)|>1$. This comes again from the linearization problem affecting all estimators above $k_c$, but also from the $k$ factor in the Legendre transform (\ref{eqesti2}), which magnifies the error on $\ha_\m$ following (\ref{eqerri1}).

Similar results will be obtained for other distributions which, as for the Gaussian, are unbounded for $x>0$ and $x<0$. In this case, $|k_c^{\pm}|$ will grow with $\m$, as in the Gaussian case, with a speed given by the tail behavior of the distribution considered, following our results of Sec.~\ref{secres}.

\subsection{One-sided exponential distributions}

\begin{figure*}[t]
\centering
\resizebox{\textwidth}{!}{\includegraphics{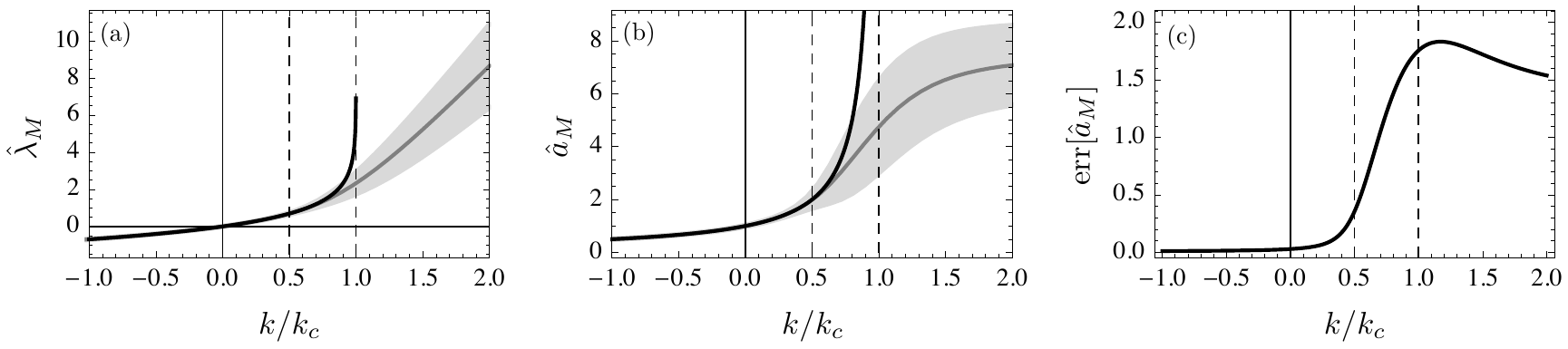}}
\caption{SCGF estimators for exponential random variables. Parameters: $\mu=1$, $\m=1000$, $\r=200$, $k_c=k_c^+=1$. (a) Gray curve: $\hlambda_\m(k)$ with statistical error (gray shaded area). Black curve: Exact $\lambda(k)$. (b) Gray curve: $\ha_\m(k)$ with statistical error (gray shaded area). Black curve: Exact derivative of $\lambda(k)$. (c) Statistical error of $a_\m(k)$.}
\label{figexp1}
\end{figure*}

We consider next the exponential distribution
\be
p(x)=\frac{1}{\mu} \E^{-x/\mu},\qquad x\geq 0
\ee
with mean $\mu$, as representative of random variables that are bounded from below. This distribution corresponds to the limit case $\rho=1$ in (\ref{eqexpdist1}) and therefore falls in principle outside the results of \cite{arous2005}. Given however that its SCGF
\be
\lambda(k)=-\ln (1-\mu k)
\ee
is defined for $k<1/\mu$, we can set $k_c^+=1/\mu$, since $\hlambda_\m(k)$ is defined for all $k\in\reals$, whereas $\lambda(k)$ exists only for $k<1/\mu$, so that the part of $\hlambda_\m(k)$ beyond $k_c^+$ is a finite sample artifact. This constant $k_c^+$ is also consistent with our estimate (\ref{eqkcest1}) of $k_c(\m)$ and arises for any distributions with exact or asymptotic exponential tails. On the other hand, we find $k_c^-=-\infty$, since $p(x)$ is bounded below at $x_{\min}=0$, so that the minimum of the sample $\{X^{(j)}\}_{j=1}^\m$ converges rapidly to 0. 

Figure~\ref{figexp1}(a) shows the result of $\hlambda_\m(k)$ for a sample size $\m=1000$ and statistical errors calculated with $\r=200$ repetitions, plotted as a function of $k/k_c$. The linearization effect is clearly seen for $\hlambda_\m(k)$, as well as for $\ha_\m(k)$, which correctly saturates to the lower bound $x=0$ for $k\ra -\infty$, but incorrectly saturates for $k>k_c^+$; see Fig.~\ref{figexp1}(b). The main difference with the Gaussian case is that, since $k_c^+$ is now constant, the convergence of $\hlambda_\m(k)$ to $\lambda(k)$  is not accompanied by an increased region of $k$ where this convergence takes place; all that changes as $\m\ra\infty$ is the slope of $\hlambda_\m$ or, equivalently, the value $\ha_\m(k)$, which diverges to reach the asymptote of $\lambda(k)$. Because $k_c^-=-\infty$, we also see that the repetition error of both $\hlambda_\m(k)$ and $\ha_\m(k)$ converges uniformly to 0 for all $k<k_c^+/2$. In Figs.~\ref{figexp1}(a) and \ref{figexp1}(b), the error bars in that region are actually smaller than the width of the black lines representing $\lambda(k)$ and $a_\m(k)$, respectively. For $k>k_c^+/2$, the error is similar to the Gaussian case: it sharply increases between $k_c^+/2$ and $k_c^+$ and saturates for $k>k_c^+$, providing again a way to estimate $k_c$.

\begin{figure}[t]
\centering
\includegraphics{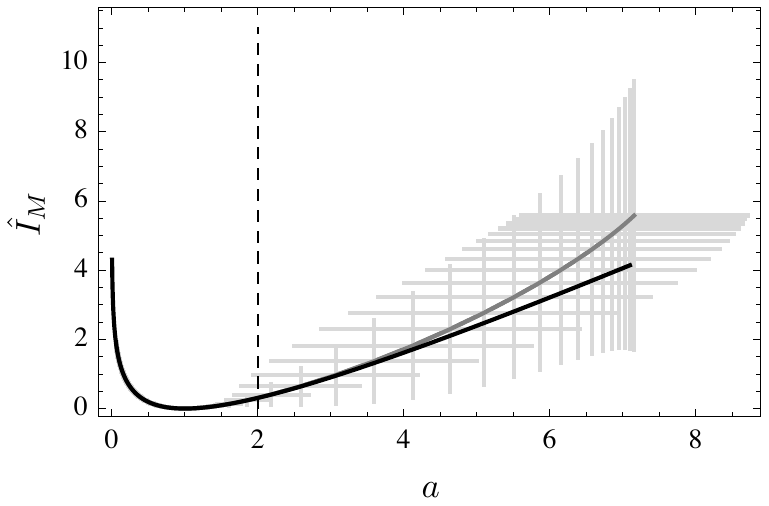}
\caption{Rate function estimator for exponential random variables. Gray curve: $\hI_\m(a)$ with statistical error bars. Black curve: Exact result. Dashed line: $a(k)$ for $k_c^+/2=1/(2\mu)$. Parameters: $\mu=1$, $\m=1000$, $\r=200$. }
\label{figexp2}
\end{figure}

The asymmetric convergence of $\hlambda_\m$ and $\ha_\m$ leads naturally to an asymmetry in the estimation of the rate function, shown in Fig.~\ref{figexp2}. For $a<a(k_c^+/2)$ the rate function is correctly estimated and matches the exact rate function
\be
I(a)=\frac{a}{\mu}-1-\ln\frac{a}{\mu},\qquad a>0,
\ee
with errors bars smaller than the width of the curve representing this function, whereas for $a>a(k_c^+/2)$, the linearized $\hlambda_\m$ and $\ha_\m$ lead to an estimation of $I(a)$ with very large error bars. 

Other distributions with asymptotic exponential tails lead to similar results. In particular, for distributions with left and right exponential tails, $k_c^-$ and $k_c^+$ are both constant with $\m$.

\subsection{Bounded distributions}

\begin{figure*}[t]
\centering
\resizebox{\textwidth}{!}{\includegraphics{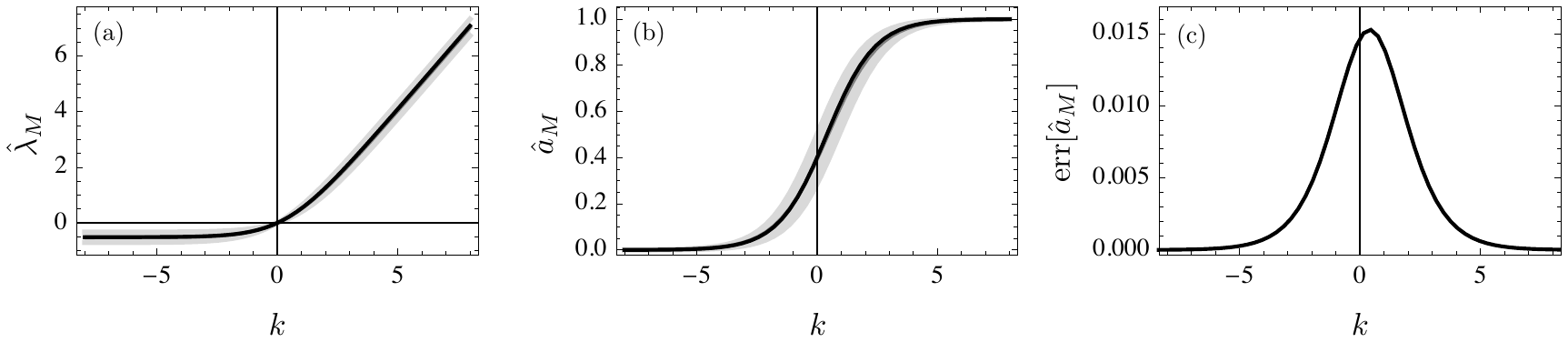}}
\caption{SCGF estimators for Bernoulli random variables. Parameters: $\alpha=0.4$, $\m=20$, $\r=1$. (a) Gray curve: $\hlambda_\m(k)$ with statistical error (gray shaded area). Black curve: Exact $\lambda(k)$. (b) Gray curve: $\ha_\m(k)$ with its statistical error (gray shaded area). Black curve: Exact derivative of $\lambda(k)$. (c) Statistical error of $a_\m(k)$.}
\label{figbern1}
\end{figure*}

The last test case of interest is the class of bounded random variables for which $\lambda(k)$ is exactly or asymptotically linear and so for which $k_c^-=-\infty$ and $k_c^+=\infty$. To illustrate this case, we consider Bernoulli random variables taking values $\{0,1\}$ with  probabilities $p(1)=\alpha$ and $p(0)=1-\alpha$, where $\alpha\in[0,1]$, so that
\be
\lambda(k)=\ln (\alpha\, \E^{k}+1-\alpha),\qquad k\in\reals.
\ee

\begin{figure}[t]
\centering
\includegraphics{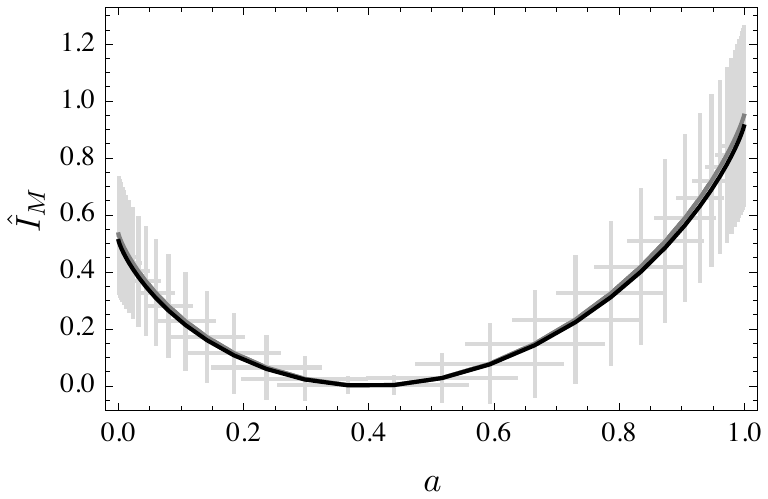}
\caption{Rate function estimator for Bernoulli random variables. Gray curve: $\hI_\m(a)$ with statistical error bars. Black curve: Exact result. Parameters: $\alpha=0.4$, $\m=20$, $\r=1$.}
\label{figbern2}
\end{figure}

Figure~\ref{figbern1} shows the results of estimating this SCGF obtained for $\m=20$ and plotted now as a function of $k$ and not $k/k_c$, since $k_c=\infty$. Already for such a small sample size, the estimators $\hlambda_\m$ and $\ha_\m$ are very accurate, compared to $\m=1000$ used in the Gaussian and exponential cases. The single and repetition errors essentially agree for all $k$ and decrease uniformly for all $k$ as $\m\ra\infty$. Figure~\ref{figbern1}(c) shows more precisely that the statistical error of $\ha_\m(k)$ is largest at $k=0$ and decreases to 0 as $k\ra\pm \infty$. This is due to the fact that the `true' linear behavior of $\lambda(k)$ as $k\ra\pm\infty$ is correctly estimated as soon as the sample $\{X^{(j)}\}$ contains one $0$ and one $1$, whereas the exact form of $\lambda(k)$ around $k=0$ depends on $\alpha$, which is correctly estimated as $\m\ra\infty$. However, both regions have errors of the same magnitude, which translate into uniform errors for the estimated rate function, shown in Fig.~\ref{figbern2}. Here we see that, despite the small sample size $\m=20$, the estimator is close to the exact rate function
\be
I(a)=a\ln\frac{a}{\alpha}+(1-a)\ln\frac{1-a}{1-\alpha},\qquad a\in[0,1],
\ee
with error bars that are significantly reduced if we were to use $\m=1000$. This comes again from the fact that linearization is not an artifact in this case: the bounded support of $p(x)$ is covered by the sample for a finite $\m$, which means essentially that $k_c=\infty$.

These results confirm previous results obtained for data networks \cite{crosby1997,duffield1995,lewis1998,duffy2005}, showing that the estimation of large deviation functions from a data stream of bits converge fast and uniformly. For other distributions with bounded support, convergence is expected to be as fast as for the Bernoulli case, with the difference that $k_c$ may not be equal to $\infty$ following our results (\ref{eqpest1}) and (\ref{eqkcest2}). For a distribution $p(x)$ that vanishes linearly, for example, we obtain $k_c\sim \m^{1/2}$ from (\ref{eqkcest1}), whereas if $p(x)$ decays like a parabola, we obtain $k_c\sim \m^{1/3}$. 

Distributions that have a fixed `window' or `vertical cut-off', such as the uniform distribution or the Bernoulli distribution, represent a limit case of bounded distributions for which $k_c=\infty$. These distributions behave similarly, whether they are discrete or continuous, because their SCGFs have asymptotic linear branches, which is the property responsible for $k_c=\infty$.

\subsection{Divergent generating functions}

To close this section, we briefly discuss the case where the theoretical SCGF diverges everywhere except at $k=0$, which arises when $X$ is distributed for example according to a L\'evy stable distribution or, more generally, any power-law tail distributions \cite{touchette2009}. Assuming that the distribution of interest is two-sided with both tails behaving as a power-law, then $\lambda(k)=\infty$ for $k\neq 0$, which is consistent with the fact that the distribution of $A_n$ does not have a large deviation form; see Example 4.2 of \cite{touchette2009}. 

The estimator $\hlambda_M(k)$ in this case exists for all $k$ when computed on finite samples, since it is a finite sum of exponentials. It is easy to check, however, that it diverges for all $k\neq 0$ as the sample size $M$ is increased. Thus, there is no convergence region for the SCGF estimator, except at $k=0$ where we trivially have $\hlambda_M(0)=\lambda(k)=0$ for all $M$, so that $k_c$ is effectively equal to $0$. This is similar to the exponential case for which $k_c$ is set to the limit of the convergence region of $\lambda(k)$. Here we have  $k_c^-=k_c^+=0$ simply because the convergence region of a distribution with left and right power-law tails is $\{0\}$. If the distribution has only one power-law tail, say to the right, then $k_c^+=0$, whereas $k_c^-$ will behave according to the shape of the other tail following the cases above.

This applies if there is no bound (experimental or numerical) on the values of $X$ that can be measured. If we increase $M$ without increasing measurement bounds, then $\hlambda_\m(k)$ will of course behave as if the quantity sampled is bounded, and will thus represent the distribution of that bounded quantity having a finite SCGF rather than the theoretical unbounded quantity having an infinite SCGF. One could then argue that all physical quantities are bounded because of the limited range of measuring devices. However, this is not a fundamental limit: measurement bounds can always be pushed in principle with better devices. From this point of view, a quantity should be considered as unbounded when the theoretical range of values that can be measured can always be made large enough to include the maximum and minimum values actually measured in any given large but finite samples.

\section{Correlated observables}
\label{secdisc}

We assumed in the previous sections that the $X_i$'s were independent to illustrate in the simplest way possible the linearization effect limiting the estimation of large deviation functions. We now consider observables involving correlated random variables, representing, for example, the individual state of interacting particles or the state of a stochastic process sampled in time. In many cases of interest, these observables involve weakly interacting components (in space or time) which can be grouped into independent or asymptotically-independent blocks. This is the basis of the \emph{block averaging} method, proposed independently in the context of free energy calculations \cite{zuckerman2002b} and large deviation theory \cite{crosby1997,duffield1995,lewis1998,duffy2005}.

We explain this method following \cite{duffy2005}; see also \cite{touchette2011}. We consider again the sample mean
\be
A_n=\frac{1}{n}\sum_{i=1}^n X_i,
\ee
but assume now that the sequence of random variables $X_1,X_2,\ldots,X_n$ forms a Markov chain. In this case, the SCGF of $A_n$ defined in (\ref{eqgenscgf1}) does not simplify to the simple cumulant function (\ref{simplescgf1}). However, it is possible to `group' the $X_i$'s into blocks as
\begin{multline}
\underbrace{X_1+\cdots +X_b}_{Y_1} + \underbrace{X_{b+1}+\cdots+X_{2b}}_{Y_2}+\cdots\\
+\underbrace{X_{n-b+1}+\cdots+X_n}_{Y_K},
\end{multline}
where $K=n/b$ is the number of blocks of size $b$, so as to rewrite the sample mean as
\be
A_n=\frac{1}{bK}\sum_{i=1}^K Y_i.
\ee
For \emph{mixing} Markov chains having a finite correlation length, it can be shown that the blocks $Y_i$ become independent in the limit where $n\ra\infty$ and $b\ra\infty$ but with $b$ growing slower than $n$ so that $K\ra\infty$ \cite{duffy2005}. Moreover, if the chain is ergodic, then the $Y_i$'s become identically distributed for $i$ large enough, so that
\be
\frac{1}{n}\ln\langle e^{nkA_n}\rangle\approx \frac{K}{n}\ln\langle\E^{k Y_i}\rangle=\frac{1}{b}\ln\langle\E^{k Y_i}\rangle.
\ee
We are thus back to the problem of estimating the SCGF for an IID sequence of random variables formed by the $Y_i$'s instead of the $X_i$'s, so that our estimator for $\lambda(k)$ is now
\be
\hlambda_{M}(k)=\frac{1}{b}\ln \frac{1}{M}\sum_{j=1}^M \E^{k Y^{(j)}},
\ee
where $Y^{(j)}$, $j=1,\ldots, M$ are $M$ IID copies of the blocked random variable $Y$. The estimation of the rate function follows as before from (\ref{eqesti1}).

In practice, this block averaging method works well by choosing a finite $b$ greater than the correlation length of the Markov chain or, equivalently, by varying $b$ until the estimated SCGF of $A_{n}$ ceases to depend on $b$. This avoids taking the double limit $n\ra\infty$, $b\ra\infty$ with $K=n/b\ra\infty$. 


The method can also be used for integrated observables of continuous-time Markov processes having the form
\be
A_T=\frac{1}{T}\int_0^T f(X_t)dt,
\label{eqaddobs1}
\ee
where $f$ is an arbitrary function of the state $X_t$ of the Markov process. In this case, the integral is `blocked' in segments of length $b$ to obtain
\be
A_T=\frac{1}{bK}\sum_{i=1}^K Y_{i},
\label{eqaddobs2}
\ee
where $Y_{i}$ is the integral of $f(X_t)$ over the time interval $[(i-1)b,ib]$ and $K=T/b$ is as before the number of blocks over the total interval $[0,T]$.

Other observables that can be expressed in the block form (\ref{eqaddobs2}) include the total activity of interacting particle systems, defined as the total number of particle jumps accumulated over a time $T$, the total integrated current which depends on the jumps and their direction, and observables of equilibrium systems. For example, one can divide the energy $E_N$ of an $N$-particle system into additive blocks $Y_{i}$ involving $b<N$ particles which become asymptotically decoupled as the limits $N\ra\infty$ and $b\ra\infty$ are taken, with $b$ growing slower than $N$. In this limit, $E_N$ is thus effectively treated as a sample mean of $K=N/b$ IID random variables. This works so long as the interactions between particles are weak or short-range, which is the spatial analog of a mixing Markov process.

In all cases, the distribution of the IID or near-IID blocks $Y_i$ determines how quickly the estimated SCGF converges to its theoretical value according to the test cases studied previously. If this distribution has an unbounded support and decays faster than an exponential, then the convergence threshold $k_c$ of the SCGF is expected to grow slowly with the sample size $M$, as in the Gaussian case, whereas if its tails decay like an exponential, then $k_c$ is essentially constant, as seen before. Bounded block distributions, on the other hand, are characterized by a $k_c$ that grows rapidly with $M$, and represent the best possible case in terms of estimation. 

The use of block averaging techniques is important as it yields an exponential gain in estimation compared to the direct sampling of large deviation probabilities. To see this, suppose that we want to estimate the probability $p$ of an event known to scale in a large deviation way as $p\sim \E^{-n I}$ with the parameter $n$, which can be $N$ or $T$ as above. It is known (see \cite{bucklew2004} or \cite{asmussen2007}) that the direct sampling of that probability requires roughly $\m\sim \E^{n}$ samples to obtain a relative error
\be
r_\m=\frac{\hP_\m-p}{p} 
\ee
for the estimate $\hP_\m$ of $p$ that is constant in $n$. By contrast, the estimation of $p$ via $\hlambda_\m$ and $\hI_\m$ leads from our results to an error on the actual rate of decay $I$ that decreases with $\m$ as $1/\sqrt{\m}$ in the convergence region. As this error is multiplied in the large deviation form of $p$ by $n$, we must therefore choose $\m>n^2$ to obtain a constant relative error for $\hP_\m=\E^{-n\hI_\m}$ as a function of $n$. 

This exponential sampling gain ($\E^n$ vs $n^2$) can obviously be exploited if $A_n$ can be divided into independent or asymptotically independent blocks for a large enough block size $b$. If this cannot be done or if $A_n$ does not have an additive structure, then we can still obtain the rate function of $A_n$ in principle by directly sampling its generating function
$
G_n(k)=\Esp{\E^{nkA_n}}
$
and obtaining its SCGF using (\ref{eqgenscgf1}). However, in this case the estimation is inefficient: the saddle-point $a^*$ of $G_{n}(k)$, which does not scale with $n$, can be reached only with a sample size $\m\sim \E^{n}$ because of the exponential form of $P(A_n=a)$. 

Considering our result (\ref{eqkcest1}), this means that $\m$ must grow exponentially with $n$ for $k_c$ to remain constant as $n$ is increased. Since $\m\sim\E^n$ is also, as just mentioned, the sample size needed to obtain the rate function of $A_n$ by direct sampling, we see that the generating function method offers no real gain over direct sampling when $A_n$ has no obvious additive structure \cite{touchette2011}. Similar results were obtained in the context of free energy estimation \cite{jarzynski2006}, where $A_n$ is the work performed on an $N$-particle system over a time $T$ so that $n=NT$, and for multifractals \cite{angeletti2011}, where $A_n$ is the local dimension measured over a spatial or temporal scale $\epsilon=1/n$. 

For experiments, there is no obvious way to overcome this problem of sampling observables that are not additive; however, for simulations, faster convergence can be achieved using modified sampling techniques, such as importance sampling \cite{touchette2011}, escort distributions~\cite{minh2006,minh2008,vaikuntanathan2008}, and transition path sampling \cite{dellago2009}, which modify the distribution of $A_n$ to center it essentially at the saddle-point $a^*(k)$. Cloning techniques \cite{lecomte2007a,tailleur2007b,giardina2011}, which are not based on sampling but rather attempt to obtain $\lambda(k)$ from the multiplicative property of generating functions, can also be used and prove efficient in simulations.

\vspace*{0.2in}
\section{Conclusion}
\label{seccon}

We have developed in this paper a general method for estimating large deviation functions from simulation or experimental data and have provided convergence results for estimators of these functions and their errors. Our results establish a separation between bounded random variables, for which the estimation of large deviation functions converges quickly as a function of sample size, and unbounded random variables, for which convergence is guaranteed only for a certain parameter region, which depends on the tail of the distribution considered. We have proposed a way to determine this convergence region without the a priori knowledge of that distribution, based on the fact that statistical errors behave differently inside and outside of that region, and have illustrated our approach for various distributions of interest.

These results can be applied to compute rate functions of any additive observables of equilibrium, nonequilibrium, and manmade systems, in addition to computing multifractal spectra, dispersion exponents, and equilibrium free energies using the Jarzynski estimator, as these are also based on estimating generating functions. Our focus on large deviations brings a new and general insight into these computations, which should play an important role in future experiments designed to probe the fluctuations of microscopic and mesoscopic systems.

\begin{acknowledgments}
H.T.\ thanks Ken Duffy and Tom\'as Tangarife for useful discussions, and the Galileo Galilei Institute for Theoretical Physics and INFN for hospitality and support during the workshop `Advances in Nonequilibrium Statistical Mechanics'. We also thank a referee for useful comments on a previous version of the paper. We gratefully acknowledge financial support to C.R.\ (Postdoctoral Programme of the Vice Rector for Research, Stellenbosch University), to F.A.\ (NITheP Postdoctoral Fellowship) and to H.T.\ (Stellenbosch University project funding for new appointee). 
\end{acknowledgments}
\bibliography{masterbib} 
 
\end{document}